# Tailoring Interdigitated Back Contacts for High-performance Bifacial Silicon Solar Cells

Yubo Sun,[1] Zhiguang Zhou,[1,2] Reza Asadpour,[1] Muhammad A. Alam[1] and Peter Bermel[1,2,a]

[1]*Department of Electrical and Computer Engineering, Purdue University, West Lafayette, IN 47907, USA*

[2]*Birck Nanotechnology Center, Purdue University, 1205 W. State St., West Lafayette, IN 47907, USA*

Photovoltaic (PV) cells have become one of the most promising renewable energy technologies. To make PV more competitive with incumbent technologies, higher power output densities are needed. One promising approach is to add bifaciality to existing monofacial PV devices, allowing more output power from the additional reflection of sunlight from the ground (albedo $\alpha$). For example, bifaciality can be added to Silicon Heterojunction (SHJ) solar cell with Interdigitated Back Contacts (IBC) by opening up the gaps between the back metal contacts, but the optimum gap ($w$) that maximizes power output is unknown. In this paper we show that that the optimum gap ($w = 1 - (1+\alpha)(\alpha(\alpha+c))^{-\frac{1}{2}}$) maximizes IBC-SHJ bifacial power output ($P \propto (1+\alpha)(1 - 2\sqrt{a/c})$), where $c$ is the ratio of output power density to power loss due to shadowing and Joule heating, The results are validated by self-consistent finite-element device modeling. For a typical α = 0.3, an optimized bifacial IBC SHJ cell will produce 17% more power output than state-of-the-art monofacial IBC SHJ cells. The results encourage development of bifacial IBC solar cells as a next generation PV technology.

The energy output of a solar cell depends on both the solar irradiance absorbed via the photovoltaic effect, as well as its efficiency in converting photo-generated carriers into electricity. Traditional *monofacial* cells accept light only from the front surface; therefore, reflection from index mismatch and/or a front-contact metallic grid reduces light coupling into cells. Texturing the front surface and/or including an anti-reflection coating addresses the challenge of index mismatch, while interdigitated back-contact cells (IBC) move both n- and p-type contacts to the back to minimize front-grid light reflection. This improved light coupling allows IBC cells to achieve particularly high power output under typical illumination. By inserting high bandgap ($E_g$) material between absorber and highly recombination active metal contacts, Silicon heterojunction solar cells (SHJ)[1] can provide higher efficiency ($\eta > 25\%$) and lower temperature coefficients[2] than traditional p-n junction solar cells. Today, despite the fact that recent market forecasts predict the rapid increase in *monofacial* Passivated Emitter and Rear Contact (PERC) cell production[3,4], *monofacial* IBC SHJ cells ranks among the very best high-performance cells, with efficiencies exceeding 26% in experiments[5].

A new solar cell architecture, called *bifacial* PV, has recently emerged as a promising technological pathway to higher output yields and lower costs[4,6–9]. The *bifacial* design accepts light from both surfaces, therefore it allows absorption of ground reflected sunlight (albedo) at the rear side of solar cell. The sum of the direct sunlight and the albedo illumination increases photo-current within the cell. Unfortunately, the benefits of *bifacial* operation of an IBC SHJ cell are not easily quantified – after all, the dense interdigitated grid in the back may only allow a fraction of the albedo (α) light to reach the cell and the excess photo-current may be lost as Joule heating in the fingers and busbars. Very little previous work has focused on assessing *bifacial* IBC SHJ cells, which are expected to be optically superior to *bifacial* front and back contact PERC cells at any α.

[a] Electronic mail: pbermel@purdue.edu.

In this letter, we address the fundamental question of *bifacial* gain of an IBC SHJ cell with sophisticated numerical modeling plus a simple analytical formula capturing the essential physics, which agree closely. We show that IBC SHJ cells benefit from *bifacial* operation provided that: (a) the albedo exceeds a critical intensity, and (b) the gaps between the metal fingers are optimized.

To tackle the back contact design of *bifacial* IBC SHJ cell, we begin by specifying the details of the device structure. As a reference, we consider a high-efficiency *monofacial* IBC SHJ cell reported by Yoshikawa et al.[5] with Aluminum (Al) metal grids. Although transparent conducting oxides (TCO) are a theoretically ideal replacement, they are not considered here, since there is not yet an ideal p-type TCO electrode for solar cells[10]. Fig. 1a shows that our device has a 165 μm n-type c-Si absorber sandwiched between 10 & 20 nm thick intrinsic a-Si passivation layers at front and rear, respectively. Interdigitated p$^+$ a-Si emitter and n$^+$ a-Si back-surface field (BSF) layers are embedded in the rear a-Si passivation layer. The BSF width $W_{BSF}$ and emitter width $W_{emit}$ are taken as 100 μm and 250 μm respectively for the following reasons: (a) *monofacial* IBC cells are typically optimized at $W_{emit} \sim 2 - 3 \times W_{BSF}$[11–13]; (b) the lateral transport path (center-to-center distance between p$^+$ and n$^+$ regions) must be comparable to the absorber thicknesses to avoid excessive bulk recombination. Their spacing is set to $W_s = 5$ μm, because Nichiprouk et al.[13] has emphasized that proximity between the emitter and BSF layers improves device efficiencies. Therefore, the half pitch width $W_P$ (the lateral distance between centers of adjacent metal grids), as labeled in Fig. 1b, is 180 μm. Finally, Al metal grids (thickness of $t_{Al}$, width of $W_{Al}$ and length of $L_{Al}$ as labeled in Fig. 1a) are pasted to the p$^+$ emitter and n$^+$ BSF to form the electrodes.

The spacing (gap) between Al metal grids ($W_g$) is the key variable controlling efficiencies, see Fig. 1b. Once albedo photons transmit through the gap in the metallic back-contact, they generate electron-holes pairs within the device. These photo-generated carriers first transport through emitter/BSF layer to reach the metal contact (*intrinsic* resistance) and then have to travel through metal fingers and/or busbars to reach the electrode (*extrinsic* resistance). As a result, thicker gridlines (i.e. narrower $W_g$) block albedo and reduce photo-current, but thinner lines (i.e. wider $W_g$) increase series resistance and reduce output power. Specifically, thinner grids (wider $W_g$) reduce the grid shadowing loss ($P_{shadow}$), but increase the series resistance loss ($P_{resist}$), since $W_{Al} = W_P - W_g$. The resulting output power ($P_{out}$) of a *bifacial* IBC SHJ cell is given by:

$$P_{OUT}(w, \alpha) = P_{ideal}(\alpha) - P_{resist}(w, \alpha) - P_{shadow}(w, \alpha); \tag{1}$$

Here, the idealized *intrinsic* power-output ($P_{ideal}$) of the IBC SHJ cell is reduced by the counter-balancing resistive and shadowing losses defined by the normalized gap width: $w = W_g/W_P$. Our goal in this paper will be to calculate the normalized gap width $w_{opt}$ that maximizes $P_{out}(w, \alpha)$.

Light absorption in a *bifacial* IBC SHJ cell involves a complex interplay of absorption in the bulk and multiple reflection by the randomly textured surface. Light absorption in a bifacial IBC SHJ cell is calculated first to determine the spatial profile of photo-generated carriers. With texturing on both sides of the cell, ray-tracing calculation may involve complicated 3D simulations[14]. Fortunately, the fact that carrier generation has only weak lateral spatial variation allows us to simplify the *optical* model to 1D. The effect of a Lambertian light trapping due to random textured surface[15] can be quantified by modifying the spectral dependent optical constant of c-Si based on the empirical formula[16]. In this model, a perfect anti-reflection coating (ARC) layer is assumed. Ray tracing is utilized to compute the absorption spectrum and spatial optical generation profile within the layered structure.

Based on the spatially-resolved light absorption profile, the transport of the photo-generated electrons and holes are solved in a 2D coupled Poisson-drift-diffusion solver *Sentaurus TCAD* [17]. The electrical properties of the interfaces and bulk layers are adapted from previous work[18]. Finally, *Sentaurus* automatically accounts for the *intrinsic* resistance related to current crowding between the metal and heavily doped regions as illustrated in Fig. 1b. For the *extrinsic* resistance, the schematic of the IBC back-contacts described by Desa et al. [19] can be used to calculate the corresponding finger and busbar resistances[20,21]: In practice, busbar resistance is negligible compared to finger resistance given by:

$$R_{finger} = \frac{1}{2} \times \frac{\rho \times L_{Al}^2}{W_{Al} \times t_{Al}} \times W_P, \tag{2}$$

where $\rho$ is the Al resistivity. This *extrinsic* series resistance associated with Al back-contact is added to the 2D *Sentaurus* device model as an external series resistance.

Our simulation framework is validated by benchmarking against experimental results of a state-of-the-art *monofacial* IBC SHJ cell[5]. Fig. 2 shows that it closely agrees with experimentally observed J-V curves and external quantum efficiency (EQE) spectra. Two non-idealities explain the remaining, small deviations observed: our non-inclusion of the imperfect ARC explains slightly higher EQE between the wavelengths of 300-400 nm, and our neglect of the series resistance explains the slightly higher fill factor (FF). The corrections are small (especially because the back-contact grids can be much wider in their *monofacial* counterparts): the overall discrepancy (in terms of power production) is below 0.5%.

Our simulation framework can now be used to optimize the design of the *bifacial* IBC SHJ cell shown in Fig. 1. A number of IBC SHJ solar cells with different $W_g$ are created in *Sentaurus* and illuminated by various albedo $\alpha$. Next, following the optoelectronic procedure described above, the output power $P_{out}(w, \alpha)$ is calculated. Here, we assume typical finger parameters $t_{Al} = 40$ μm thick[22,23] and $L_{Al} = 10$ cm long[5,19]; $W_{Al} = W_P - W_g = W_P \cdot (1 - w)$ is our parameter for performance optimization of *bifacial* IBC SHJ solar cells. To quantify shadowing and resistive losses, the results are post-processed, such that the resistive losses due to Al fingers can be extracted from the difference between power output with and without the *extrinsic* resistance.

By definition[24], the power output $P_{OUT}$ of solar cell is given by:

$$P_{OUT} = V_{OC} \times J_{SC} \times FF, \tag{3}$$

where $J_{SC}$ is the short circuit current; $V_{OC}$ is the open circuit voltage; and FF is the fill factor. These four key metrics ($V_{OC}$, $J_{SC}$, FF and $P_{OUT}$) for assessing solar cell performance are plotted as a function of $w$ in Fig. 3. $J_{SC}$ is dictated by total amount of sunlight received from front and rear side. It increases linearly with increasing $w$ (see Fig. 3a), as the gap between the electrodes allows larger fraction of the albedo light to enter the cell and contribute to photo-generation. The loss of FF in Fig. 3b is a consequence of increasing series resistance with increment of $w$. Increasing $w$ impacts the series resistance by (1) reducing the overlap region between metal and emitter/BSF layer to aggravate the current crowding effect; (2) narrowing $W_{Al}$ to increase $R_{finger}$ according to Eq. (1). Since $V_{OC}$ scales logarithmically[24] with $J_{SC}$, the increase in $V_{OC}$ in Fig. 3c is easily explained. $V_{OC}$ is relatively insensitive to the variation of $w$, because $V_{OC}$, as a metric that measures how good carrier transport is within the absorber, is driven by the diffusion length of minority carriers, length of quasi-neutral region and surface defect density etc.[25], none of which is directly correlated to $w$ between Al fingers. Since $J_{SC}$ and $FF$ respond inversely to $w$, $P_{OUT}$ as a function of $w$ shows a concave downward curve, as shown in Fig. 3d. For increased α, the optimum $w$ that delivers the maximum $P_{OUT}$ shifts to larger values, as summarized in Table I:

Table I. Optimum $w$ to produce maximum power output with given α

| α | 0.1 | 0.3 | 0.5 | 0.7 | 0.9 |
|---|---|---|---|---|---|
| $w$ | 0.67 | 0.72 | 0.78 | 0.78 | 0.78 |
| Max. $P_{OUT}$ (mW/cm²) | 27.7 | 31.3 | 35.0 | 38.8 | 42.3 |

At a practical albedo of α = 0.3, total resistive loss and the optical loss of shadowing limit the realistic maximum $P_{OUT}$ at 31.3 mW/cm² with a $w_{opt}$ of 0.72, which exceeds $P_{OUT}$ of *monofacial* IBC SHJ cell by 4.5 mW/cm² (17% relative increase). Although the self-shading effect in the monolithic module[26] has not been explicitly accounted for in this work, it can be captured by reducing α. In this case, the improvement of $P_{OUT}$ from *monofacial* IBC SHJ cell to *bifacial* IBC SHJ cells is still substantial.

The numerical optimization above was performed for a particular IBC SHJ Al-interconnected cell. Different groups may design *bifacial* IBC cells variously, with different grid periodicity, contact and metal resistances, etc. Therefore a general solution for $w_{opt}(\alpha)$ is desired. To develop this generalization, we first carefully assess the performance of *bifacial* IBC SHJ cell with α = 0.3 by breaking down the loss mechanisms as depicted in Fig. 4a. Optical loss of shadowing increased linearly with the decrease of $w$ as it reflects the fact that total amount of absorbed photons scaled linearly with unshaded area. The effective contact resistive loss due to current crowding effect outweighs the finger resistive loss, regardless of $w$. Remarkably, a generalized analytical expression captures these essential features of power output by realizing $J_{sc}(w, \alpha) \cong J_{sc}(\alpha = 0) \cdot (1 + \alpha \cdot w)$. Therefore, Eq. (2) can be rewritten as:

$$P_{OUT}(w) \cong P_{OUT,mono} \cdot (1 + \alpha \cdot w) - \frac{1}{2} \cdot (L_{Al})^2 \cdot J_{MP,mono}^2 \cdot (1 + \alpha \cdot w)^2 \cdot \frac{\rho}{(1-w) \cdot t_{Al}} \quad (4)$$
$$- J_{MP,mono}^2 \cdot (1 + \alpha \cdot w)^2 \cdot \frac{\rho_c}{(1-w)};$$

in which $P_{OUT,mono}$ is the standard output power of a *monofacial* IBC SHJ solar cell; $J_{MP,mono}$ is the operating current of *monofacial* IBC SHJ cell at maximum power point; and $\rho_c$ is the contact resistivity between $Al$ grids and heavily doped regions. To find the normalized gap width $w$ that maximizes $P_{OUT}$, we set $\left.\frac{dP_{OUT}}{dw}\right|_{w=w_{opt}} = 0$; the result for $w_{opt}$ is given by:

$$w_{opt} \equiv \frac{W_g}{W_P} = 1 - (1 + \alpha)\big(\alpha(\alpha + c)\big)^{-\frac{1}{2}} \quad (5a)$$

$$c \equiv \frac{P_{OUT,mono}}{J_{MP,mono}^2 \cdot \left(\frac{1}{2} \times \frac{L_{Al}^2 \cdot \rho}{t_{Al}} + \rho_c\right)} \quad (5b)$$

where $c$ captures the ratio of the output power density for the *monofacial* IBC SHJ cell to the total resistive power loss density. Typically, this should greatly exceed unity in a good cell. At $\alpha = 0.3$, the analytic solution is benchmarked against simulation results. Strong agreement is observed in Fig. 4b. For various α, the analytically-calculated $w$ matches with the numerical simulation within discretization error. By inserting Eq. 5a & 5b into Eq. 4, the benefit of *bifacial* cell over *monofacial* cell can be quantified as a function of α and $c$, and further simplified for (α ≪ c) as follows:

$$\frac{P_{OUT}}{P_{OUT,mono}} = \frac{1+\alpha}{c} \cdot \left[c + 2\alpha - 2\sqrt{\alpha(\alpha+c)}\right] \cong (1+\alpha)\left(1 - 2\sqrt{\frac{\alpha}{c}}\right) \tag{6}$$

The optimum power scales linearly with albedo (i.e. $1 + \alpha$), although the geometric mean of albedo and the relative loss coefficient (i.e., $\sqrt{\alpha/c}$) suppresses some optical gain. The generalized analytic expression allows one to simply estimate the optimum $w$ and resulting benefit of a *bifacial* cell over a *monofacial* cell in the best case.

Interestingly, Eq. (4) also offers an insight not immediately apparent from our numerical simulations. By definition, $w \geq 0$, therefore Eq. (4) suggests that the albedo must exceed a critical value ($\alpha > \alpha_c$) to ensure that *bifacial* operation actually produced more power than its *monofacial* counterpart. By replacing α by $\alpha_c$, and setting $w_{opt} = 0$, we find that $\alpha_c = \frac{1}{c-2}$. Recall that for high quality solar cells, resistive power loss is a small fraction of the total power output, therefore $c$ is much greater than 2 and $\alpha_c$ turns out to be very close to zero. For this specific numerical example, $w_{opt} = 0$ is reached when $\alpha_c \leq 0.01$.

In summary, we have explored the optimization of back contacted design for *bifacial* IBC SHJ solar cell which theoretically outperforms the fast growing PERC technique as a pathway towards low cost of PV technologies. By incorporating a realistic series resistance model that reflects both *extrinsic* and *intrinsic* resistive losses into our well-calibrated 2D electro-optically coupled device simulation framework, we have identified the $w_{opt}$ between Al back contacts for *bifacial* IBC SHJ solar cell under various α. For a practical α of 0.3, we find that a *bifacial* IBC SHJ cell outperforms *monofacial* IBC SHJ cell by 4.5 mW/cm² in the best case. Furthermore, a generalized analytical expression addressing the balance of resistive loss and optical loss of shadowing was also derived, which shows that bifacial IBC solar cells offer a significant advantage in power production across a broad range of albedos and materials. As a result, this study may serve as a general guideline for the design and optimization of future *bifacial* IBC SHJ solar cells.

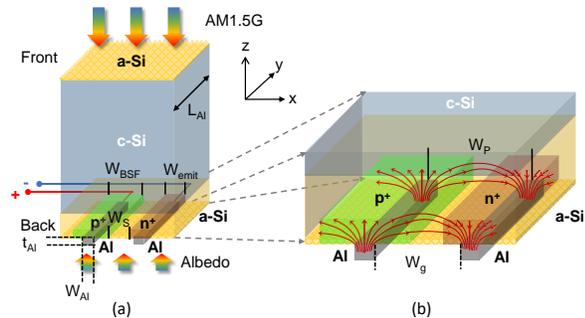

Fig 1. (a) 3D schematics of the bifacial IBC-HIT solar cell design studied in this work. Key length scales are defined here. (b) Current crowding is induced near the contacts by the narrow overlap between the emitter/BSF and the Al metal grid.

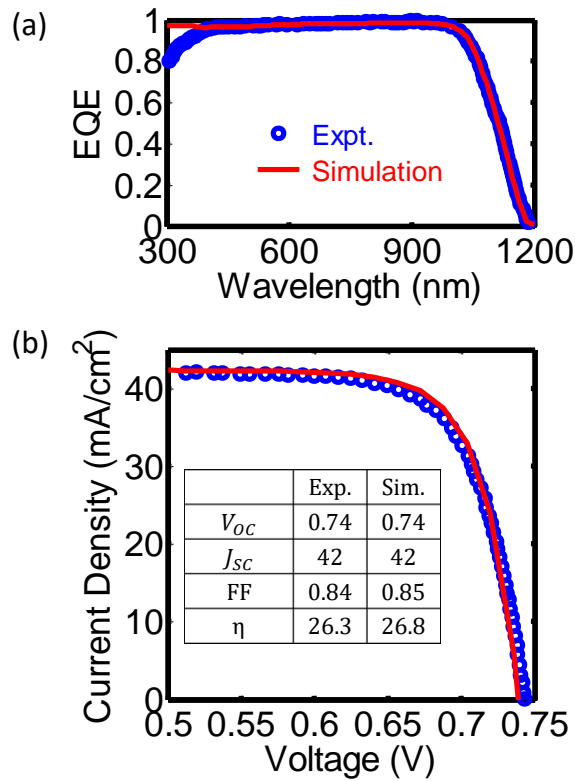

Fig 2. Our 2-D simulation framework for IBC SHJ cells incorporating analytic Lambertian light trapping shows strong agreement with (a) spectral dependent EQE results (except for $\lambda < 400$ nm ) and (b) measured I-V characteristics adapted from Yoshikawa et al.[3] (dots: experiment; line: simulation).

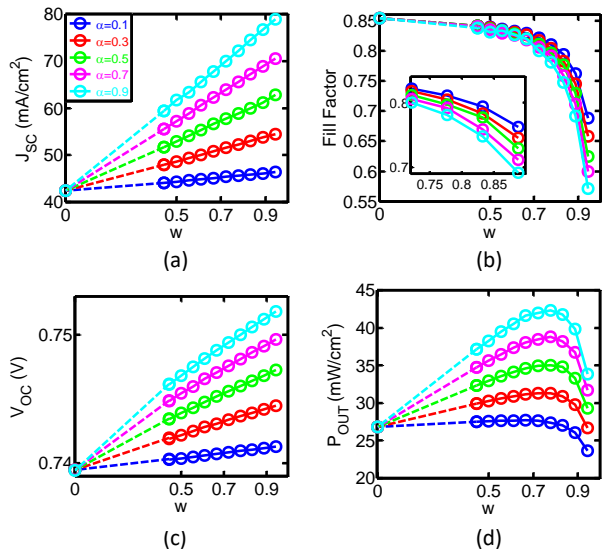

Fig 3. Performance of the bifacial SHJ cell structure as a function of reduced gap width w between Al metal contacts under various $\alpha$. (a) short circuit current; (b) fill factor; (c) open circuit voltage; (d) cell power output. (dots: numerical results; dashed line: trend lines).

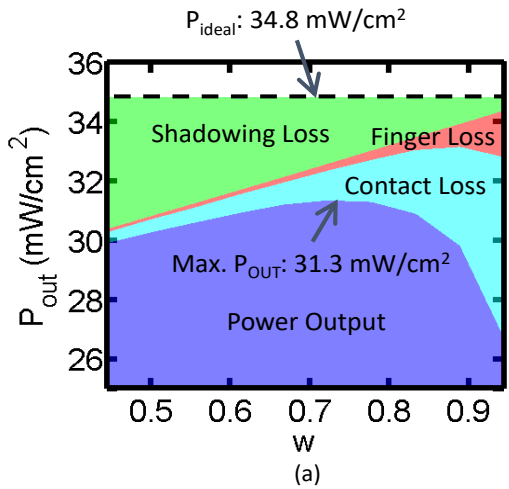

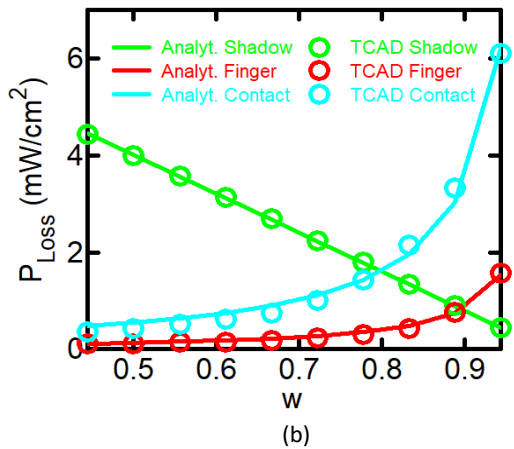

Fig 4. (a) Illustration of different loss mechanisms (optical loss of shadowing, effective contact resistance loss, finger resistance loss) under various $w$. (b) Our analytical model (lines) shows strong agreement with simulations (circles) for $\alpha = 0.3$.